\documentclass[letterpaper, 12pt]{article}[2000/05/19]
\usepackage[english]{babel}
\usepackage{amsfonts,amsmath,amssymb,amsthm,latexsym,amscd,mathrsfs}
\usepackage{ifthen,cite}
\usepackage[bookmarksnumbered=true]{hyperref}
\hypersetup{pdfpagetransition={Split}}

\newcommand{\evenhead}{Author \ name}
\newcommand{\oddhead}{Article \ name}
\newcommand{\theArticleName}{Article \ name}

\newcommand{\FirstPageHeading}[1]{\thispagestyle{empty}%
\noindent\raisebox{0pt}[0pt][0pt]{\makebox[\textwidth]{\protect\footnotesize \sf }}\par}

\newcommand{\ArticleName}[1]{\renewcommand{\theArticleName}{#1}\vspace{-2mm}\par\noindent {\LARGE\bf  #1\par}}
\newcommand{\Author}[1]{\vspace{5mm}\par\noindent {\Large  #1\par} \par\vspace{2mm}\par}
\newcommand{\Address}[1]{\vspace{2mm}\par\noindent {\it #1} \par}
\newcommand{\Email}[1]{\ifthenelse{\equal{#1}{}}{}{\par\noindent {\rm E-mail: }{\it  #1} \par}}
\newcommand{\URLaddress}[1]{\ifthenelse{\equal{#1}{}}{}{\par\noindent {\rm URL: }{\tt  #1} \par}}
\newcommand{\EmailD}[1]{\ifthenelse{\equal{#1}{}}{}{\par\noindent {$\phantom{\dag}$~\rm E-mail: }{\it  #1} \par}}
\newcommand{\URLaddressD}[1]{\ifthenelse{\equal{#1}{}}{}{\par\noindent {$\phantom{\dag}$~\rm URL: }{\tt  #1} \par}}

\newcommand{\Abstract}[1]{\vspace{6mm}\par\noindent\hspace*{10mm}
\parbox{140mm}{\small {\bf Abstract.} #1}\par}
\newcommand{\Keywords}[1]{\vspace{3mm}\par\noindent\hspace*{10mm}
\parbox{140mm}{\small {\bf Key words:} \rm #1}\par}
\newcommand{\Classification}[1]{\vspace{3mm}\par\noindent\hspace*{10mm}
\parbox{140mm}{\small {\it 2000 Mathematics Subject Classification:} \rm #1}\vspace{3mm}\par}
\newcommand{\ShortArticleName}[1]{\renewcommand{\oddhead}{#1}}
\newcommand{\AuthorNameForHeading}[1]{\renewcommand{\evenhead}{#1}}

\long\def\@makecaption#1#2{
  \sbox\@tempboxa{\small \textbf{#1.}\ \ #2}%
  \ifdim \wd\@tempboxa >\hsize
    {\small \textbf{#1.}\ \ #2}\par \else
    \global \@minipagefalse
    \hb@xt@\hsize{\hfil\box\@tempboxa\hfil}%
  \fi \vskip\belowcaptionskip}

\def\numberwithin#1#2{\@ifundefined{c@#1}{\@nocounterr{#1}}{%
  \@ifundefined{c@#2}{\@nocnterr{#2}}{%
  \@addtoreset{#1}{#2}%
  \toks@\@xp\@xp\@xp{\csname the#1\endcsname}%
  \@xp\xdef\csname the#1\endcsname
    {\@xp\@nx\csname the#2\endcsname.\the\toks@}}}}
\def\E^#1{{\buildrel #1 \over\vee}}
\newtheorem{theorem}{Theorem}

{\theoremstyle{definition}

}

\begin{document}

\FirstPageHeading{V.I. Gerasimenko}

\ShortArticleName{Quantum kinetic evolution of observables}

\AuthorNameForHeading{V.I. Gerasimenko}

\ArticleName{Quantum Kinetic Evolution\\ of Marginal Observables}

\Author{V.I. Gerasimenko\footnote{E-mail: \emph{gerasym@imath.kiev.ua}}}

\Address{Institute of Mathematics of NAS of Ukraine,\\
    3, Tereshchenkivs'ka Str.,\\
    01601, Kyiv-4, Ukraine}

\bigskip

\Abstract{We develop a rigorous formalism for the description
of the evolution of observables of quantum systems of particles in the mean-field scaling limit.
The corresponding asymptotics of a solution of the initial-value
problem of the dual quantum BBGKY hierarchy is constructed. Moreover, links of the evolution
of marginal observables and the evolution of quantum states described in terms of
a one-particle marginal density operator are established. Such approach gives the alternative
description of the kinetic evolution of quantum many-particle systems to generally accepted approach
on basis of kinetic equations.}

\bigskip


\vspace{2pc}
\Classification{35Q40; 47d06.}

\makeatletter
\renewcommand{\@evenhead}{
\hspace*{-3pt}\raisebox{-15pt}[\headheight][0pt]{\vbox{\hbox to \textwidth {\thepage \hfil \evenhead}\vskip4pt \hrule}}}
\renewcommand{\@oddhead}{
\hspace*{-3pt}\raisebox{-15pt}[\headheight][0pt]{\vbox{\hbox to \textwidth {\oddhead \hfil \thepage}\vskip4pt\hrule}}}
\renewcommand{\@evenfoot}{}
\renewcommand{\@oddfoot}{}
\makeatother

\newpage
\vphantom{math}

\protect\tableofcontents

\vspace{0.7cm}

\section{Introduction}

During the last decade, the considerable advance in the rigorous derivation of quantum kinetic equations,
in particular the nonlinear Schr\"{o}dinger equation and the Gross-Pitaevskii equation
\cite{AGT,AA,BGGM2,ESchY2,FL,MKRM,PP09} and the quantum Boltzmann equation \cite{BCEP3,ESY}, is observed.

It is well known that a description of quantum many-particle systems is formulated in
terms of two sets of objects: observables and states. The functional of the mean value of observables
defines a duality between observables and states and as a consequence there exist two approaches
to the description of evolution. Usually the evolution of many-particle systems is described
in the framework of the evolution of states by the quantum BBGKY hierarchy for marginal density operators.
An equivalent approach of the description of the evolution of quantum systems is given in terms of marginal
observables by the dual quantum BBGKY hierarchy (the Heisenberg picture of evolution) \cite{BG,BGer}.

The conventional philosophy of the description of kinetic evolution consists in the following.
The evolution of states can be effectively described by a one-particle marginal density operator governed by
the kinetic equation as a result of some approximations \cite{BQ} or in suitable scaling limits \cite{Sh}.
In the paper we develop an approach to the description of kinetic evolution of quantum many-particle systems
in framework of the evolution of marginal observables.
For this purpose we investigate the mean-field asymptotics of a solution of the initial-value problem
of the quantum dual BBGKY hierarchy. In addition links between the evolution
of observables and the kinetic evolution of states described in terms of a one-particle marginal
density operator are discussed in the general case.

We now outline the structure of the paper and the main results.
In Section 2 we introduce some preliminary definitions and construct a solution
of the Cauchy problem of the dual quantum BBGKY hierarchy for marginal observables
as an expansion over particle clusters which evolution is governed by the corresponding-order
cumulant (semi-invariant) of the groups of operators of the Heisenberg equations of finitely many particles.
In Section 3 the main result is proved, namely, the mean-field asymptotic behavior
of stated above solution of the dual quantum BBGKY hierarchy is established. The constructed
asymptotics is governed by the recurrence evolution equations set (the dual quantum Vlasov hierarchy).
In Section 4 typical properties of the dual kinetic dynamics are formulated, in particular
the relation of the dual quantum Vlasov hierarchy with the nonlinear Schr\"{o}dinger equation is considered.
Moreover, the relation of the dual quantum BBGKY hierarchy and the generalized quantum kinetic equation is established.
Finally we conclude with some observations and perspectives for future research.

\section{The evolution of observables of quantum many-particle systems}

In order to construct the asymptotic form of the marginal ($s$-particle) observables of quantum many-particle systems
we describe the evolution by means of the dual quantum BBGKY hierarchy.
We introduce such a hierarchy of evolution equations
and formulate necessary properties of a solution of the Cauchy problem of this hierarchy.

\subsection{Preliminary facts}

We consider a quantum system of a non-fixed, i.e. arbitrary but finite, number of identical spinless
particles obeying Maxwell-Boltzmann statistics in the space $\mathbb{R}^{\nu},$ $\nu\geq 1$. We will use units where
$h={2\pi\hbar}=1$ is a Planck constant,  and $m=1$ is the mass of particles.
The Hamiltonian $H={\bigoplus\limits}_{n=0}^{\infty}H_{n}$ of such system is a self-adjoint operator with the domain
$\mathcal{D}(H)=\{\psi=\oplus\psi_{n}\in{\mathcal{F}_{\mathcal{H}}}\mid \psi_{n}\in\mathcal{D}
(H_{n})\in\mathcal{H}_{n},\,{\sum\limits}_{n}\|H_{n}\psi_{n}\|^{2}<\infty\}\subset{\mathcal{F}_{\mathcal{H}}}$, where
$\mathcal{F}_{\mathcal{H}}={\bigoplus\limits}_{n=0}^{\infty}\mathcal{H}^{\otimes n}$ is the Fock space over the Hilbert space
$\mathcal{H}$. We adopt the usual convention that $\mathcal{H}^{\otimes 0}=\mathbb{C}$.
Assume $\mathcal{H}=L^{2}(\mathbb{R}^\nu)$ then an element $\psi\in\mathcal{F}_{\mathcal{H}}
={\bigoplus\limits}_{n=0}^{\infty}L^{2}(\mathbb{R}^{\nu n})$ is a sequence of functions
$\psi=\big(\psi_0,\psi_{1}(q_1),\ldots,\psi_{n}(q_1,\ldots,q_{n}),\ldots\big)$ such that
$\|\psi\|^{2}=|\psi_0|^{2}+\sum_{n=1}^{\infty}\int dq_1\ldots dq_{n}|\psi_{n}(q_1,\ldots ,q_{n})|^{2}<+\infty.$
On the subspace of infinitely differentiable functions with compact supports
$\psi_n\in L^{2}_0(\mathbb{R}^{\nu n})\subset L^{2}(\mathbb{R}^{\nu n})$
the Hamiltonian $H_{n}$ of $n\geq1$ particles acts according to the formula
\begin{equation}\label{H}
   H_{n}\psi_n =\sum\limits_{i=1}^{n}K(i)\psi_n +\epsilon\sum\limits_{i<j=1}^{n}\Phi(i,j)\psi_{n}.
\end{equation}
where $K(i)\psi_n = -\frac{1}{2}\Delta_{q_i}\psi_n$ is the operator of the kinetic energy,
$\Phi(i,j)\psi_{n}= \Phi(|q_{i}-q_{j}|)\psi_{n}$ is the operator of a two-body interaction
potential $\Phi$ and $\epsilon>0$ is a scaling parameter. Hereinafter we shall consider the bounded interaction potentials.

Let a sequence $g=(g_{0},g_{1},\ldots,g_{n},\ldots)$ be an infinite sequence
of self-adjoint bounded operators $g_{n}$ defined on the Fock space
$\mathcal{F}_{\mathcal{H}}$. An operator $g_{n}$ defined on the $n$-particle
Hilbert space $\mathcal{H}_{n}=\mathcal{H}^{\otimes n}$ will be denoted by $g_{n}(1,\ldots,n)$.
Let the space $\mathfrak{L}(\mathcal{F}_\mathcal{H})$ be the space of sequences
$g=(g_{0},g_{1},\ldots,$ $g_{n},\ldots)$ of
bounded operators $g_{n}$ defined on the Hilbert space
$\mathcal{H}_n$ that satisfy symmetry condition:
$ g_{n}(1,\ldots,n)=g_{n}(i_1,\ldots,i_n)$, for arbitrary $(i_{1},\ldots,i_{n})\in (1,\ldots,n)$,
equipped with the operator norm $\|.\|_{\mathfrak{L}(\mathcal{H}_{n})}$ \cite{DauL}.
We will also consider a more general space $\mathfrak{L}_{\gamma}(\mathcal{F}_\mathcal{H})$ with a norm
\begin{equation*}
   \big\|g\big\|_{\mathfrak{L}_{\gamma} (\mathcal{F}_\mathcal{H})}\doteq
   \max\limits_{n\geq 0}\, \frac{\gamma^n}{n!}\,\big\|g_{n}\big\|_{\mathfrak{L}(\mathcal{H}_{n})},
\end{equation*}
where $0<\gamma<1$. We denote by $\mathfrak{L}_{_{\gamma},0}(\mathcal{F}_\mathcal{H})
\subset\mathfrak{L}_{\gamma}(\mathcal{F}_\mathcal{H})$ the everywhere dense set in
of finite sequences of degenerate operators with infinitely differentiable kernels with compact supports.

Observables of finitely many quantum particles are sequences of self-adjoint
operators from the space $\mathfrak{L}_{\gamma}(\mathcal{F}_\mathcal{H})$.
The case of unbounded operators of observables can be reduced to the case under consideration \cite{DauL}.

Let $\mathfrak{L}^{1}(\mathcal{F}_\mathcal{H})= {\bigoplus\limits}_{n=0}^{\infty}
\mathfrak{L}^{1}(\mathcal{H}_{n})$ be the space of sequences
$f=(I,f_{1},\ldots,f_{n},\ldots)$ of trace class operators
$f_{n}=f_{n}(1,\ldots,n)\in\mathfrak{L}^{1}(\mathcal{H}_{n})$, satisfying the mentioned above symmetry condition,
equipped with the trace norm
\begin{equation*}
      \big\|f\big\|_{\mathfrak{L}^{1}(\mathcal{F}_\mathcal{H})}=
      \sum\limits_{n=0}^{\infty}\,\big \|f_{n}\big\|_{\mathfrak{L}^{1}(\mathcal{H}_{n})}\doteq
      \sum\limits_{n=0}^{\infty}\,\mathrm{Tr}_{1,\ldots,n}|f_{n}(1,\ldots,n)|,
\end{equation*}
where $\mathrm{Tr}_{1,\ldots,n}$ is the partial trace over $1,\ldots,n$ particles.
The everywhere dense set of finite sequences of degenerate operators with infinitely differentiable
kernels with compact supports in the space $\mathfrak{L}^{1}(\mathcal{F}_\mathcal{H})$
we denote by $\mathfrak{L}^{1}_0(\mathcal{F}_\mathcal{H})$.

The sequences of operators $f_{n}\in\mathfrak{L}^{1}(\mathcal{H}_{n}),$ $n\geq 1$, which kernels are known
as density matrices defined on the $n$-particle Hilbert space $\mathcal{H}_{n}=L^{2}(\mathbb{R}^{\nu n})$,
describe the states of a quantum system of non-fixed number of particles.

The space $\mathfrak{L}(\mathcal{F}_\mathcal{H})$ is dual to the space $\mathfrak{L}^{1}(\mathcal{F}_\mathcal{H})$
with respect to the bilinear form
\begin{equation}\label{averageD}
        \langle g|f\rangle\doteq\sum\limits_{n=0}^{\infty}\frac{1}{n!}\,\mathrm{Tr}_{1,\ldots,n}\,g_{n}f_{n},
\end{equation}
where $g_{n}\in\mathfrak{L}(\mathcal{H}_{n})$ and $f_{n}\in\mathfrak{L}^{1}(\mathcal{H}_{n})$.
The mean value of observables are given by the positive
continuous linear functional \eqref{averageD} on the space of observables $\mathfrak{L}_{\gamma}(\mathcal{F}_\mathcal{H})$.

\subsection{The dual quantum BBGKY hierarchy}

The evolution of marginal observables is described by the initial-value problem
of the dual quantum BBGKY hierarchy
\begin{equation}\label{dh}
\begin{split}
   \frac{d}{dt}G_{s}(t,Y)=&\big(\sum\limits_{i=1}^{s}\mathcal{N}_{0}(i)+
     \epsilon\sum\limits_{i<j=1}^{s}\mathcal{N}_{\mathrm{int}}(i,j)\big)G_{s}(t,Y)\\
   & +\epsilon\sum_{j_1\neq j_{2}=1}^s
     \mathcal{N}_{\mathrm{int}}(j_1,j_{2})G_{s-1}(t,Y\backslash (j_1)),
\end{split}
\end{equation}
\begin{equation}\label{dhi}
   \hskip-50mm G_{s}(t)\mid_{t=0}=G_{s}^0,\quad s\geq1,
\end{equation}
where on $\mathfrak{L}_{0}(\mathcal{H}_n)\subset\mathfrak{L}(\mathcal{H}_n)$
the operators $\mathcal{N}_{0}$ and $\mathcal{N}_{\mathrm{int}}$ are consequently defined by formulas
\begin{equation}\label{com}
\begin{split}
   & \mathcal{N}_{0}(j)g_n\doteq -i \big[g_n,K(i)\big],\\
   & \mathcal{N}_{\mathrm{int}}(i,j)g_n\doteq -i\big[g_n,\Phi(i,j)\big],
\end{split}
\end{equation}
where $\big[ \cdot,\cdot \big]$ is a commutator of operators.
We refer to equations (\ref{dh}) as the dual quantum BBGKY hierarchy since the canonical quantum BBGKY hierarchy \cite{CGP97,Pe95,GerS}
for the marginal density operators is the dual hierarchy of evolution equations with respect to bilinear form (\ref{averageD})
to evolution equations (\ref{dh}).
In case of the space $\mathcal{H}=L^{2}(\mathbb{R}^\nu)$, evolution equations (\ref{dh})
for kernels of the operators $G_{s}(t)$, $s\geq 1$, are given by
\begin{equation*}
\begin{split}
    i\,\frac{\partial}{\partial t}G_{s}&(t,q_1,\ldots,q_s;q'_1,\ldots,q'_s)=
       \big(-\frac{1}{2}\sum\limits_{i=1}^s(-\Delta_{q_i}+\Delta_{q'_i})\\
    &\hskip+7mm +\epsilon\sum\limits_{1=i<j}^s\big(\Phi(q'_i-q'_j)-\Phi(q_i-q_j)\big)\big)
       G_s(t,q_1,\ldots,q_s;q'_1,\ldots,q'_s)\\
    &\hskip+12mm +\epsilon\sum\limits_{1=i\neq j}^s\big(\Phi(q'_i-q'_j)
       -\Phi(q_i-q_j)\big)G_{s-1}(t,q_1,\ldots,q_{j-1},q_{j+1},\ldots,q_s;\\
    & \hskip+77mm q'_1,\ldots,q'_{j-1},q'_{j+1},\ldots,q'_s), \quad s\geq1.
\end{split}
\end{equation*}

To construct a solution of abstract initial-value problem (\ref{dh})
we formulate some necessary facts. For $g_n\in\mathfrak{L}(\mathcal{H}_{n})$ it is defined the one-parameter mapping
\begin{equation}\label{grG}
    \mathbb{R}^1\ni t\mapsto\mathcal{G}_n(t)g_n\doteq e^{itH_{n}}g_n e^{-itH_{n}}.
\end{equation}
On the space $\mathfrak{L}(\mathcal{H}_{n})$ one-parameter mapping (\ref{grG})
is an isometric $\ast$-weak continuous group of operators, i.e. it is a $C_{0}^{\ast}$-group.
The infinitesimal generator $\mathcal{N}_{n}$ of this group of operators is a closed operator for the $\ast$-weak topology
and on its domain of the definition $\mathcal{D}(\mathcal{N}_{n})\subset\mathfrak{L}(\mathcal{H}_{n})$
it is defined in the sense of the $\ast$-weak convergence of the space $\mathfrak{L}(\mathcal{H}_{n})$ by the operator \cite{BR}
\begin{equation}\label{infOper1}
    \mathrm{w^{\ast}-}\lim\limits_{t\rightarrow 0}\frac{1}{t}\big(\mathcal{G}_n(t)g_n-g_n \big)
    =-i(g_n H_n-H_n g_n)\doteq\mathcal{N}_n g_n,
\end{equation}
where  $H_{n}$ is the Hamiltonian (\ref{H})
and the operator: $\mathcal{N}_n g_n=-i(g_n H_n-H_n g_n)$ is defined on the domain $\mathcal{D}(H_n)\subset\mathcal{H}_n.$

Let us introduce some abridged notations: $Y\equiv(1,\ldots,s)$, $X\equiv (j_1,\ldots,j_{n})\subset Y$,
and $\{Y\backslash X\}$ is the set consisting of one element $X=(j_1,\ldots,j_{n})$,
i.e. the set $(j_1,\ldots,j_{n})$ is a connected subset of the partition $\mathrm{P}$ such that
$|\mathrm{P}|=1$ ($|\mathrm{P}|$ denotes the number of partitions of a set).

We define the $n$-order cumulant of group of operators (\ref{grG}) as follows \cite{G,GerRS}
\begin{equation}\label{cumulant}
    \mathfrak{A}_{1+n}\big(t,\{Y\backslash X\}, X\big)\doteq
    \sum\limits_{\mathrm{P}:\,(\{Y\backslash X\}, X)={\bigcup}_i X_i}
    (-1)^{\mathrm{|P|}-1}({\mathrm{|P|}-1})!\prod_{X_i\subset \mathrm{P}}\mathcal{G}_{|X_i|}(t,X_i),\quad n\geq0,
\end{equation}
where ${\sum}_\mathrm{P}$ is the sum over all possible partitions $\mathrm{P}$
of the set $(\{Y\backslash X\},j_1,\ldots,j_{n})$ into
$|\mathrm{P}|$ nonempty mutually disjoint subsets  $ X_i\subset (\{Y\backslash X\}, X)$.

Let us indicate some properties of cumulants (\ref{cumulant}).
If $n=1$, for $g_{1}\in\mathfrak{L}_0(\mathcal{H})\subset\mathfrak{L}(\mathcal{H})$
in the sense of the $\ast$-weak convergence of the space $\mathfrak{L}(\mathcal{H})$
the generator of the first-order cumulant is given by the operator
\begin{equation*}
    \lim\limits_{t\rightarrow 0}\frac{1}{t}(\mathfrak{A}_{1}(t,1)-I) g_{1}(1)=\mathcal{N}_{0}(1)g_{1}(1).
\end{equation*}
In the case $n=2$ for $g_{2}\in\mathfrak{L}_0(\mathcal{H}_{2})\subset\mathfrak{L}(\mathcal{H}_{2})$
we have in the sense of the $\ast$-weak convergence of the space $\mathfrak{L}(\mathcal{H}_{2})$
\begin{equation*}
    \lim\limits_{t\rightarrow 0}\frac{1}{t}\,\mathfrak{A}_{2}(t,1,2) g_{2}(1,2)
    =\epsilon\,\mathcal{N}_{\mathrm{int}}(1,2)g_{2}(1,2),
\end{equation*}
and for $n>2$ as a consequence of the fact that we consider a system of particles interacting
by a two-body potential, it holds
\begin{equation*}
    \lim\limits_{t\rightarrow 0}\frac{1}{t}\,\mathfrak{A}_{n}(t,1,\ldots,n) g_{n}(1,\ldots,n)=0.
\end{equation*}

On the space $\mathfrak{L}_{\gamma}(\mathcal{F}_\mathcal{H})$ for abstract initial-value problem (\ref{dh})-(\ref{dhi})
the following statement holds \cite{BG}.

The solution $G(t)=(G_{0},G_{1}(t,1),\ldots,G_{s}(t,Y),\ldots)$ of the Cauchy problem (\ref{dh})-(\ref{dhi})
of the dual quantum BBGKY hierarchy is defined
provided that $\gamma<e^{-1}$ by the expansions
\begin{equation}\label{sdh}
    G_{s}(t,Y)=\sum_{n=0}^s\,\frac{1}{n!}\sum_{j_1\neq\ldots\neq j_{n}=1}^s
    \mathfrak{A}_{1+n}\big(t,\{Y\backslash X\},X\big)\,G_{s-n}(0,Y\backslash X),\quad s\geq 1,
\end{equation}
where the $(1+n)$-order cumulant $\mathfrak{A}_{1+n}\big(t,\{Y\backslash X\},X\big)$
is determined by formula (\ref{cumulant}) and the estimate holds
\begin{equation}\label{es}
  \big\|G(t)\big\|_{\mathfrak{L}_{\gamma}(\mathcal{F}_\mathcal{H})}
  \leq e^2(1-\gamma e)^{-1}\big\|G(0)\big\|_{\mathfrak{L}_{\gamma}(\mathcal{F}_\mathcal{H})}.
\end{equation}
For $G(0)\in\mathfrak{L}_{\gamma}^0(\mathcal{F}_\mathcal{H})\subset\mathfrak{L}_{\gamma}(\mathcal{F}_\mathcal{H})$
it is a classical solution and for arbitrary initial data $G(0)\in\mathfrak{L}_{\gamma}(\mathcal{F}_\mathcal{H})$
it is a generalized solution.

We note that expansion (\ref{sdh}) can be represented in the form of the perturbation (iteration) series
as a result of applying of analogs of the Duhamel equation
to cumulants (\ref{cumulant}) of groups of operators (\ref{grG}) from solution expansion (\ref{sdh}) \cite{GerS}.

As stated above (see formula (\ref{averageD})) the mean value of the marginal observable
$G(t)\in\mathfrak{L}_{\gamma}(\mathcal{F}_\mathcal{H})$ at $t\in \mathbb{R}$
in the initial marginal state $F(0)=(I,F_{1}(0,1),\ldots,F_{s}(0,Y),\ldots)\in\mathfrak{L}^{1}(\mathcal{F}_\mathcal{H})$
is defined by the functional
\begin{equation}\label{avmar-1}
       \big\langle G(t)\big|F(0)\big\rangle=\sum\limits_{s=0}^{\infty}\,\frac{1}{s!}\,
       \mathrm{Tr}_{\mathrm{1,\ldots,s}}\,G_{s}(t,1,\ldots,s)F_{s}(0,1,\ldots,s).
\end{equation}
According to estimate (\ref{es}), functional (\ref{avmar-1}) exists under the condition that $\gamma<e^{-1}$.

\section{The dual quantum Vlasov hierarchy}

We consider the problem of the rigorous description of the quantum kinetic evolution on the basis of
many-particle dynamics of observables by way of example of the mean-field asymptotic behavior
of stated above solution of the dual quantum BBGKY hierarchy.

\subsection{A mean-field limit of the dual quantum BBGKY hierarchy}

Consider the mean-field scaling limit of a solution of initial-value problem (\ref{dh})-(\ref{dhi})
of the dual quantum BBGKY hierarchy.
\begin{theorem}\label{3.1} If for initial data $G_{s}(0)\in\mathfrak{L}(\mathcal{H}_{s})$
there exists the limit $g_{s}(0)\in\mathfrak{L}(\mathcal{H}_{s})$
\begin{equation}\label{asumdin}
    \mathrm{w^{\ast}-}\lim\limits_{\epsilon\rightarrow 0}\big( \epsilon^{-s} G_{s}(0)-g_{s}(0)\big)=0,
\end{equation}
then for arbitrary finite time interval there exists the mean-field limit of solution (\ref{sdh})
of the Cauchy problem (\ref{dh})-(\ref{dhi}) of the dual quantum BBGKY hierarchy in the sense of
the $\ast$-weak convergence of the space $\mathfrak{L}(\mathcal{H}_s)$
\begin{equation}\label{asymt}
    \mathrm{w^{\ast}-}\lim\limits_{\epsilon\rightarrow 0} \big(\epsilon^{-s} G_{s}(t)-g_{s}(t)\big)=0,
\end{equation}
and it is defined by the expansion
\begin{equation}\label{Iterd}
\begin{split}
   & g_{s}(t,Y)=\sum\limits_{n=0}^{s-1}\,\int\limits_0^tdt_{1}\ldots\int\limits_0^{t_{n-1}}dt_{n}
      \, \mathcal{G}_{s}^{0}(t-t_{1})\sum\limits_{i_{1}\neq j_{1}=1}^{s}
      \mathcal{N}_{\mathrm{int}}(i_{1},j_{1})\,\mathcal{G}_{s-1}^{0}(t_{1}-t_{2})\\
   & \hskip+25mm \ldots\,\, \mathcal{G}_{s-n+1}^{0}(t_{n-1}-t_{n})
      \sum\limits^{s}_{\mbox{\scriptsize $\begin{array}{c}i_{n}\neq j_{n}=1,\\
      i_{n},j_{n}\neq (j_{1},\ldots,j_{n-1})\end{array}$}}
      \mathcal{N}_{\mathrm{int}}(i_{n},j_{n})\\
   & \hskip+50mm \times \mathcal{G}_{s-n}^{0}(t_{n})g_{s-n}(0,Y\backslash (j_{1},\ldots,j_{n})),\quad s\geq1,
\end{split}
\end{equation}
where the following notation of the group of operators (\ref{grG}) of noninteracting particles is used
\begin{equation*}
\begin{split}
 \mathcal{G}_{s-n+1}^{0}(t_{n-1}-t_{n})
       &\equiv\mathcal{G}_{s-n+1}^{0}(t_{n-1}-t_{n},Y \backslash (j_{1},\ldots,j_{n-1}))\\
   &=\prod\limits_{j\in Y \backslash (j_{1},\ldots,j_{n-1})}\mathcal{G}_{1}(t_{n-1}-t_{n},j).
\end{split}
\end{equation*}
\end{theorem}

Before to prove this statement we give some comments.
If $g(0)\in\mathfrak{L}(\mathcal{F}_\mathcal{H})$, the sequence $g(t)=(g_0,g_1(t),\ldots,$ $g_{s}(t),\ldots)$
of limit marginal observables (\ref{Iterd}) is a generalized global solution of the initial-value problem of
the \emph{dual quantum Vlasov hierarchy}
\begin{equation}\label{vdh}
    \frac{d}{dt}g_{s}(t,Y)=\sum\limits_{i=1}^{s}\mathcal{N}_{0}(i)\,g_{s}(t,Y)+
    \sum_{j_1\neq j_{2}=1}^s\mathcal{N}_{\mathrm{int}}(j_1,j_{2})\,g_{s-1}(t,Y\backslash(j_1)),
\end{equation}
\begin{equation}\label{vdhi}
   \hskip-75mm g_{s}(t)\mid_{t=0}=g_{s}^0,\quad s\geq1.
\end{equation}
This fact is proved similar to the case of an iteration series of the dual quantum BBGKY hierarchy \cite{BG}.
It should be noted that equations set (\ref{vdh}) has the structure of recurrence evolution equations.
We make a few examples of the dual quantum Vlasov hierarchy (\ref{vdh}) in terms of operator kernels of
the limit marginal observables
\begin{equation*}
\begin{split}
    &i\,\frac{\partial}{\partial t}g_{1}(t,q_1;q'_1)=-\frac{1}{2}(-\Delta_{q_1}+\Delta_{q'_1})g_{1}(t,q_1;q'_1),\\
    &i\,\frac{\partial}{\partial t}g_{2}(t,q_1,q_2;q'_1,q'_2)=
       \big(-\frac{1}{2}\sum\limits_{i=1}^2(-\Delta_{q_i}+\Delta_{q'_i})+(\Phi(q'_1-q'_2)-\Phi(q_1-q_2))\big)\\
    &\hskip+21mm\times g_{2}(t,q_1,q_2;q'_1,q'_2)+\big(\Phi(q'_1-q'_2)-
       \Phi(q_1-q_2)\big)\big(g_{1}(t,q_1;q'_1)+g_{1}(t,q_2;q'_2)\big).
\end{split}
\end{equation*}

Let us consider a particular case of observables, namely
the mean-field limit of the additive-type marginal observables\footnote{The $k$-ary  marginal observable is the sequence $G^{(k)}(0)=\big(0,\ldots,0,G_{k}^{(k)}(0,1,\ldots,k),0,\ldots\big)$ \cite{BGer}.}:
$G^{(1)}(0)=(0,G_{1}^{(1)}(0,1),0,\ldots)$.
In this case solution (\ref{sdh}) of the dual BBGKY hierarchy (\ref{dh}) has the form
\begin{equation}\label{sad}
    G_{s}^{(1)}(t,Y)=\mathfrak{A}_{s}(t)\,\sum_{j=1}^s\,G_{1}^{(1)}(0,j),
\end{equation}
where $\mathfrak{A}_{s}(t)$ is the $s$-order cumulant (\ref{cumulant}) of groups of operators (\ref{grG}).
If for the additive-type observables $G^{(1)}(0)$ condition (\ref{asumdin}) is satisfied, i.e. it holds
\begin{equation*}
    \mathrm{w^{\ast}-}\lim\limits_{\epsilon\rightarrow 0}\big( \epsilon^{-1} G_{1}^{(1)}(0)-g_{1}^{(1)}(0)\big)=0,
\end{equation*}
then according to statement (\ref{asymt}), we have
\begin{equation*}
    \mathrm{w^{\ast}-}\lim\limits_{\epsilon\rightarrow 0} \big(\epsilon^{-s} G_{s}^{(1)}(t)-g_{s}^{(1)}(t)\big)=0,
\end{equation*}
where the limit operator $g_{s}^{(1)}(t)$ is defined by the expression
\begin{equation}\label{itvad}
\begin{split}
   & g_{s}^{(1)}(t,Y)=\int\limits_0^t dt_{1}\ldots\int\limits_0^{t_{s-2}}dt_{s-1}
     \,\mathcal{G}_{s}^{0}(t-t_{1})\sum\limits_{i_{1}\neq j_{1}=1}^{s} \mathcal{N}_{\mathrm{int}}(i_{1},j_{1})
     \,\mathcal{G}_{s-1}^{0}(t_{1}-t_{2})\\
   &\hskip+21mm \ldots \,\,\mathcal{G}_{2}^{0}(t_{s-2}-t_{s-1})
     \sum\limits^{s}_{\mbox{\scriptsize $\begin{array}{c}i_{s-1}\neq j_{s-1}=1,\\
        i_{s-1},j_{s-1}\neq (j_{1},\ldots,j_{s-2})\end{array}$}}\mathcal{N}_{\mathrm{int}}(i_{s-1},j_{s-1})\\
   &\hskip+45mm \times \mathcal{G}_{1}^{0}(t_{s-1})\,g_{1}^{(1)}(0,Y\backslash (j_{1},\ldots,j_{s-1})),\quad s\geq1,
\end{split}
\end{equation}
as a special case of expansion  (\ref{Iterd}). We give examples of expressions (\ref{itvad})
\begin{equation*}
\begin{split}
   & g_{1}^{(1)}(t,1)=\mathcal{G}_{1}(t,1)\,g_{1}^{(1)}(0,1),\\
   & g_{2}^{(1)}(t,1,2)=\int\limits_0^t d\tau\,\prod\limits_{i=1}^{2}\mathcal{G}_{1}(t-\tau,i)\,\mathcal{N}_{\mathrm{int}}(1,2)
     \sum\limits_{j=1}^{2}\mathcal{G}_{1}(\tau,j)\,g_{1}^{(1)}(0,j).
\end{split}
\end{equation*}

\subsection{Proof of Theorem \ref{3.1}}

In case of bounded interaction potential (\ref{H}) for the
group of operators (\ref{grG}) the analog of the Duhamel equation is valid
\begin{equation*}\label{iter2kum}
\begin{split}
    & \big(\mathcal{G}_{s}(t,Y)-\mathcal{G}_{s-1}(t,Y\backslash j_1)\mathcal{G}_{1}(t,j_1)\big)g_s\\
    = &\,\,\epsilon\int\limits_{0}^{t}d\tau\,\mathcal{G}_{s}(t-\tau,1,\ldots,s)
       \sum\limits^{s}_{\mbox{\scriptsize $\begin{array}{c}i_{1}=1,\\
       i_{1}\neq j_{1}\end{array}$}}\mathcal{N}_{\mathrm{int}}(i_{1},j_{1})\,
       \mathcal{G}_{s-1}(\tau,Y\backslash j_1)\mathcal{G}_{1}(\tau,j_1)g_s\\
    = &\,\,\epsilon\int\limits_{0}^{t}d\tau\,\mathcal{G}_{s-1}(t-\tau,Y\backslash j_1)\mathcal{G}_{1}(t-\tau,j_1)
       \sum\limits^{s}_{\mbox{\scriptsize $\begin{array}{c}i_{1}=1,\\
       i_{1}\neq j_{1}\end{array}$}}\mathcal{N}_{\mathrm{int}}(i_{1},j_{1})\,\mathcal{G}_{s}(\tau,1,\ldots,s)g_s,
\end{split}
\end{equation*}
where the operator $\mathcal{N}_{\mathrm{int}}(i,j)$ is defined by formula (\ref{com}) and $Y\equiv(1,\ldots,s)$.
Then for $(1+n)$-order cumulant of groups of operators (\ref{grG}) the analog of the Duhamel equation holds
\begin{equation}\label{itcum}
\begin{split}
   & \mathfrak{A}_{1+n}\big(t,\{Y\backslash (j_1,\ldots,j_{n})\},j_1,\ldots,j_{n}\big)\,
       G_{s-n}(0,Y\backslash (j_1,\ldots,j_{n}))\\
   =&\,\,\epsilon^n\, n! \int\limits_0^tdt_{1}\ldots\int\limits_0^{t_{n-1}}dt_{n}
       \, \mathcal{G}_{s}(t-t_{1})\sum\limits^{s}_{\mbox{\scriptsize $\begin{array}{c}i_{1}=1,\\
       i_{1}\neq j_{1}\end{array}$}}\mathcal{N}_{\mathrm{int}}(i_{1},j_{1})\,\mathcal{G}_{s-1}(t_{1}-t_{2})\\
   & \hskip+8mm\ldots\,\, \mathcal{G}_{s-n+1}(t_{n-1}-t_{n}) \sum\limits^{s}_{\mbox{\scriptsize $\begin{array}{c}i_{n}=1,\\
       i_{n}\neq (j_{1},\ldots,j_{n})\end{array}$}}\mathcal{N}_{\mathrm{int}}(i_{n},j_{n})
       \mathcal{G}_{s-n}(t_{n})\,G_{s-n}(0,Y\backslash (j_{1},\ldots,j_{n})),
\end{split}
\end{equation}
where accepted above notations are used, $\mathcal{G}_{s-n}(t_{n})\equiv\mathcal{G}_{s-n}(t_{n},Y\backslash (j_1,\ldots,j_{n}))$
and we take into consideration the identity: $\mathcal{G}_{n}(t,X)G_{s-n}(0,Y\backslash X)=G_{s-n}(0,Y\backslash X)$.

For arbitrary finite time interval $\ast$-weak continuous group of operators (\ref{grG}) has
the following scaling limit in the sense of the $\ast$-weak convergence
of the space $\mathfrak{L}(\mathcal{H}_s)$ \cite{BR}
\begin{equation}\label{Kato}
    \mathrm{w^{\ast}-}\lim\limits_{\epsilon\rightarrow 0}\big(\mathcal{G}_s(t)g_s-
    \prod\limits_{j=1}^{s}\mathcal{G}_{1}(t,j)g_s\big)=0.
\end{equation}

Taking into account assumption (\ref{asumdin}) and an analog of the Duhamel equation (\ref{itcum}),
in view of formula (\ref{Kato}) of an asymptotic perturbation of group (\ref{grG}) for the $n$
term of expansion (\ref{sdh}) we have
\begin{equation}\label{apc}
\begin{split}
   & \mathrm{w^{\ast}-}\lim\limits_{\epsilon\rightarrow 0}
       \Big(\epsilon^{-n}\mathfrak{A}_{1+n}\big(t,\{Y\backslash X\},j_1,\ldots,j_{n}\big)\,
       \epsilon^{-(s-n)}G_{s-n}(0,Y\backslash X)\\
   & \hskip+7mm -\, n! \int\limits_0^tdt_{1}\ldots\int\limits_0^{t_{n-1}}dt_{n}
       \, \mathcal{G}_{s}^{0}(t-t_{1})\sum\limits^{s}_{\mbox{\scriptsize $\begin{array}{c}i_{1}=1,\\
       i_{1}\neq j_{1}\end{array}$}}\mathcal{N}_{\mathrm{int}}(i_{1},j_{1})\,\mathcal{G}_{s-1}(t_{1}-t_{2})\\
   & \hskip+15mm \ldots\,\, \mathcal{G}_{s-n+1}^{0}(t_{n-1}-t_{n})
       \sum\limits^{s}_{\mbox{\scriptsize $\begin{array}{c}i_{n}=1,\\
       i_{n}\neq (j_{1},\ldots,j_{n})\end{array}$}}\mathcal{N}_{\mathrm{int}}(i_{n},j_{n})
       \mathcal{G}_{s-n}^{0}(t_{n})\,g_{s-n}(0,Y\backslash X) \Big)=0,
\end{split}
\end{equation}
where it is used the following notations: $X\equiv(j_{1},\ldots,j_{n})$ and $\mathcal{G}_{s-n+1}^{0}(t_{n-1}-t_{n})
\equiv\prod\limits_{j\in Y \backslash X}\mathcal{G}_{1}(t_{n-1}-t_{n},j)$.

As a result of equality (\ref{apc}) we establish the validity of statement (\ref{asymt}) for solution (\ref{sdh})
of the dual quantum BBGKY hierarchy (\ref{dh}).

To construct the evolution equations which satisfy limit expression (\ref{Iterd})
we differentiate over the time variable expansion (\ref{Iterd})
in the sense of pointwise convergence of the space $\mathfrak{L}(\mathcal{H}_s)$.
In view equality (\ref{infOper1}) it holds
\begin{equation*}
    \frac{d}{dt}g_{s}(t,Y)=\sum\limits_{i=1}^{s}\mathcal{N}_{0}(i)\,g_{s}(t,Y)+
      \sum_{j_1\neq j_{2}=1}^s\mathcal{N}_{\mathrm{int}}(j_1,j_{2})\sum\limits_{n=0}^{s-2}\,\int\limits_0^t dt_{2}\ldots\int\limits_0^{t_{n+1}}dt_{n+1}\,\mathcal{G}_{s-1}^{0}(t-t_{2})
\end{equation*}
\begin{equation*}
\begin{split}
\\
   & \hskip+5mm\times\sum\limits^{s}_{\mbox{\scriptsize $\begin{array}{c}i_{2}\neq j_{2}=1,\\
      i_{2},j_{2}\neq j_{1}\end{array}$}}\mathcal{N}_{\mathrm{int}}(i_{2},j_{2})\,\mathcal{G}_{s-2}^{0}(t_{2}-t_{3})\ldots\,\, \mathcal{G}_{s-n}^{0}(t_{n}-t_{n+1})\\
   & \hskip+12mm\times\sum\limits^{s}_{\mbox{\scriptsize $\begin{array}{c}i_{n+1}\neq j_{n+1}=1,\\
      i_{n+1},j_{n+1}\neq (j_{1},\ldots,j_{n})\end{array}$}}
      \mathcal{N}_{\mathrm{int}}(i_{n+1},j_{n+1})\mathcal{G}_{s-n-1}^{0}(t_{n+1})g_{s-n-1}(0,Y\backslash (j_{1},\ldots,j_{n+1})).
\end{split}
\end{equation*}
According to definition (\ref{Iterd}), the second summand in the right-hand side of this equality is  expressed in terms of
the limit marginal observable $g_{s-1}(t,Y\backslash j_1)$
and consequently, we get the dual quantum Vlasov hierarchy (\ref{vdh}).

\section{Some properties of the dual kinetic dynamics}

The links of constructed mean-field asymptotic behavior of marginal observables with
the nonlinear Schr\"{o}dinger equation are considered. Furthermore the relation between the evolution
of observables and the generalized description of the kinetic evolution of states
in terms of a one-particle marginal density operator is discussed.

\subsection{The propagation of a chaos}

Hereinafter we shall consider initial data satisfying the factorization property or a "chaos" property \cite{CGP97},
which means the lack of correlations at initial time. For a system of identical particles, obeying the
Maxwell-Boltzmann statistics, we have
\begin{equation}\label{h2}
    F(t)|_{t=0}= F^{(c)}\equiv \big(F_1^0(1),\ldots,\prod_{i=1}^s F_1^0(i),\ldots\big).
\end{equation}
The assumption about initial data is intrinsic for the kinetic
description of a gas, because in this case all possible states are
characterized only by a one-particle marginal density operator. Let
\begin{equation}\label{lh2}
    \lim\limits_{\epsilon\rightarrow 0}\big\|\, \epsilon \,F_1^0-f_1^0 \,\big\|_{\mathfrak{L}^{1}(\mathcal{H})}=0,
\end{equation}
then the limit of the initial state $F^{(c)}$ satisfies a chaos property too
\begin{equation}\label{lih2}
    f^{(c)}\equiv\big(f_1^0(1),\ldots,\prod \limits_{i=1}^{s}f_{1}^0(i),\ldots\big).
\end{equation}
If $g(t)\in\mathfrak{L}_{\gamma}(\mathcal{F}_\mathcal{H})$ and $f_1^0\in \mathfrak{L}^{1}(\mathcal{H})$, then
under the condition $\|f_1^0\|_{\mathfrak{L}^{1}(\mathcal{H})}<\gamma$,
there exists the mean-field limit of mean value functional (\ref{avmar-1}) which is determined by the expansion
\begin{equation*}
    \big\langle g(t)\big|f^{(c)}\big\rangle=\sum\limits_{s=0}^{\infty}\,\frac{1}{s!}\,
    \mathrm{Tr}_{\mathrm{1,\ldots,s}}\,g_{s}(t,1,\ldots,s)\prod \limits_{i=1}^{s} f_1^0(i).
\end{equation*}

In consequence of the following equality for the limit additive-type marginal observables (\ref{itvad})
(it is proved below in more general case)
\begin{equation}\label{avmar-2}
\begin{split}
  \big\langle g^{(1)}(t)\big|f^{(c)}\big\rangle& =\sum\limits_{s=0}^{\infty}\,\frac{1}{s!}\,
       \mathrm{Tr}_{\mathrm{1,\ldots,s}}\,g_{s}^{(1)}(t,1,\ldots,s)\prod \limits_{i=1}^{s} f_{1}^0(i)\\
  & =\mathrm{Tr}_{\mathrm{1}}\,g_{1}^{(1)}(0,1)f_{1}(t,1),
\end{split}
\end{equation}
where operator $g_{s}^{(1)}(t)$ is given by (\ref{itvad}) and $f_{1}(t,1)$ is the solution
\begin{equation}\label{viter}
\begin{split}
   f_{1}(t,1)=&\sum\limits_{n=0}^{\infty}\int\limits_0^tdt_{1}\ldots\int\limits_0^{t_{n-1}}dt_{n}\,\mathrm{Tr}_{2,\ldots,n+1}
        \prod\limits_{i_1=1}^{1}\mathcal{G}_{1}(-t+t_{1},i_1)\\
   & \times\big(-\mathcal{N}_{\mathrm{int}}(1,2)\big)\prod\limits_{j_1=1}^{2}\mathcal{G}_{1}(-t_{1}+t_{2},j_1)
        \ldots\prod\limits_{i_{n}=1}^{n}\mathcal{G}_{1}(-t_{n}+t_{n},i_{n})\\
   & \hskip+12mm \times\sum\limits_{k_{n}=1}^{n}\big(-\mathcal{N}_{\mathrm{int}}(k_{n},n+1)\big)
        \prod\limits_{j_n=1}^{n+1}\mathcal{G}_{1}(-t_{n},j_n)\prod\limits_{i=1}^{n+1}f_1^0(i)
\end{split}
\end{equation}
of the initial-value problem of the Vlasov quantum kinetic equation
\begin{equation}\label{Vlasov1}
      \frac{d}{dt}f_{1}(t,1)=-\mathcal{N}_{0}(1)f_{1}(t,1)+
      \mathrm{Tr}_{2}\big(-\mathcal{N}_{\mathrm{int}}(1,2)\big)f_{1}(t,1)f_{1}(t,2),
\end{equation}
\begin{equation}\label{Vlasovi}
     \hskip-72mm f_1(t)|_{t=0}= f_1^0,
\end{equation}
we establish that hierarchy (\ref{vdh}) for additive-type marginal observables and initial state (\ref{lh2})
describe the evolution of quantum many-particle systems as by the Vlasov quantum kinetic equation (\ref{Vlasov1}).

Indeed for bounded interaction potential (\ref{H}) series (\ref{viter}) is norm convergent on the space
$\mathfrak{L}^{1}(\mathcal{H})$ under the condition
\begin{equation*}
     t<t_0\equiv\big(2\, \|\Phi\|_{\mathfrak{L}(\mathcal{H}_{2})}\|f_1^0\|_{\mathfrak{L}^{1}(\mathcal{H})}\big)^{-1},
\end{equation*}
and hence the functional in right-hand side of equality (\ref{avmar-2}) exists.
Taking into account the validity for $f_n\in\mathfrak{L}^{1}_{0}(\mathcal{H}_n)$
in the sense of the norm convergence of the equality
\begin{equation*}
     \lim\limits_{t\rightarrow 0}\frac{1}{t}(\mathcal{G}_n(-t)f_n-f_n)=-\mathcal{N}_n f_n,
\end{equation*}
where the operator $(-\mathcal{N}_n)$ is defined by formulas (\ref{com}),
we differentiate over the time variable expression (\ref{viter})
in the sense of pointwise convergence of the space $\mathfrak{L}^{1}(\mathcal{H})$
\begin{equation}\label{dife}
\begin{split}
   \frac{d}{dt}f_{1}(t,1)=&-\mathcal{N}_{0}(1)f_{1}(t,1)+
        \mathrm{Tr}_{2}\big(-\mathcal{N}_{\mathrm{int}}(1,2)\big)\\
   & \times\sum\limits_{n=0}^{\infty}\int\limits_0^tdt_{1}\ldots\int\limits_0^{t_{n-1}}dt_{n}\,
        \mathrm{Tr}_{3,\ldots,n+2}\prod\limits_{i_1=1}^{2}\mathcal{G}_{1}(-t+t_{1},i_1)
\end{split}
\end{equation}
\begin{equation*}
\begin{split}
   & \times\sum\limits_{k_{1}=1}^{2}\big(-\mathcal{N}_{\mathrm{int}}(k_{1},3)\big)
        \prod\limits_{j_1=1}^{3}\mathcal{G}_{1}(-t_{1}+t_{2},j_1)
        \ldots\prod\limits_{i_{n}=1}^{n+1}\mathcal{G}_{1}(-t_{n}+t_{n},i_{n})\\
   & \hskip+25mm \times\sum\limits_{k_{n}=1}^{n+1}\big(-\mathcal{N}_{\mathrm{int}}(k_{n},n+2)\big)
        \prod\limits_{j_n=1}^{n+2}\mathcal{G}_{1}(-t_{n},j_n)\prod\limits_{i=1}^{n+2}f_1^0(i).
\end{split}
\end{equation*}
Using the product formula for the one-particle marginal density operator $f_{1}(t,i)$
defined by expansion (\ref{viter}) for initial data (\ref{lih2})
\begin{equation}\label{prodi}
\begin{split}
    & \prod\limits_{i=1}^{k}f_{1}(t,i)=\sum\limits_{n=0}^{\infty}\int\limits_0^t
        dt_{1}\ldots\int\limits_0^{t_{n-1}}dt_{n}\mathrm{Tr}_{k+1,\ldots,k+n}
        \prod\limits_{i_1=1}^{k}\mathcal{G}_{1}(-t+t_{1},i_1)\\
    & \hskip+7mm\times\sum\limits_{k_{1}=1}^{k}\big(-\mathcal{N}_{\mathrm{int}}(k_{1},k+1)\big)
        \prod\limits_{j_1=1}^{k+1}\mathcal{G}_{1}(-t_{1}+t_{2},j_1)\ldots
        \prod\limits_{i_n=1}^{k+n-1}\mathcal{G}_{1}(-t_{n-1}+t_{n},i_n)\\
    & \hskip+25mm\times\sum\limits_{k_{n}=1}^{k+n-1}\big(-\mathcal{N}_{\mathrm{int}}(k_{n},k+n)\big)
        \prod\limits_{j_n=1}^{k+n}\mathcal{G}_{1}(-t_{n},j_n)\prod \limits_{i=1}^{k+n} f_{1}^0(i),
\end{split}
\end{equation}
where the group property of one-parameter mapping (\ref{grG}) is applied,
we express the second summand in the right-hand side of equality (\ref{dife}) in terms of ${\prod\limits}_{i=1}^{2}f_{1}(t,i)$
and consequently, we get equation (\ref{Vlasov1}).

Correspondingly, a chaos property in the Heisenberg picture of evolution of quantum many-particle systems
is fulfil. It follows from the equality for the limit $k$-ary  marginal observables,
i.e.  $g^{(k)}(0)=(0,\ldots,g_{k}^{(k)}(0,1,\ldots,k),0,\ldots)$,
\begin{equation}\label{dchaos}
\begin{split}
    \big\langle g^{(k)}(t)\big|f^{(c)}\big\rangle&=\sum\limits_{s=0}^{\infty}\,\frac{1}{s!}\,
       \mathrm{Tr}_{\mathrm{1,\ldots,s}}\,g_{s}^{(k)}(t,1,\ldots,s) \prod \limits_{i=1}^{s} f_1^0(i)\\
    & =\frac{1}{k!}\mathrm{Tr}_{\mathrm{1,\ldots,k}}\,g_{k}^{(k)}(0,1,\ldots,k)\prod \limits_{i=1}^{k} f_{1}(t,i),\quad k\geq2,
\end{split}
\end{equation}
where the limit one-particle marginal density operator $f_{1}(t,i)$ is defined by expansion (\ref{viter})
and therefore it is governed by the Cauchy problem (\ref{Vlasov1})-(\ref{Vlasovi}).

Really, taking into account the validity of the following equality
for the expression $g_{s}^{(k)}(t)$ defined by formula (\ref{Iterd})
\begin{equation*}
\begin{split}
    & \sum\limits_{s=0}^{\infty}\,\frac{1}{s!}\,\mathrm{Tr}_{\mathrm{1,\ldots,s}}\,g_{s}^{(k)}(t)
       \prod\limits_{i=1}^{s} f_1^0(i)=\frac{1}{k!}\mathrm{Tr}_{\mathrm{1,\ldots,k}}\,g_{k}^{(k)}(0)
       \sum\limits_{n=0}^{\infty}\int\limits_0^t dt_{1}\ldots\int\limits_0^{t_{n-1}}dt_{n}\\
    & \hskip+5mm\times\mathrm{Tr}_{k+1,\ldots,k+n}
        \prod\limits_{i_1=1}^{k}\mathcal{G}_{1}(-t+t_{1},i_1)
        \sum\limits_{k_{1}=1}^{k}\big(-\mathcal{N}_{\mathrm{int}}(k_{1},k+1)\big)
        \prod\limits_{j_1=1}^{k+1}\mathcal{G}_{1}(-t_{1}+t_{2},j_1)\\
    & \hskip+12mm\ldots \prod\limits_{i_n=1}^{k+n-1}\mathcal{G}_{1}(-t_{n-1}+t_{n},i_n)
        \sum\limits_{k_{n}=1}^{k+n-1}\big(-\mathcal{N}_{\mathrm{int}}(k_{n},k+n)\big)
        \prod\limits_{j_n=1}^{k+n}\mathcal{G}_{1}(-t_{n},j_n)\prod \limits_{i=1}^{k+n} f_{1}^0(i)
\end{split}
\end{equation*}
and product formula (\ref{prodi}) for $f_{1}(t,i)$ defined by series (\ref{viter}),
we finally get equality (\ref{dchaos}).

Thus, in the mean-field scaling limit an equivalent approach to the description of the kinetic evolution
of quantum many-particle systems in terms of the Cauchy problem (\ref{Vlasov1})-(\ref{Vlasovi})
of the Vlasov kinetic equation is given by the Cauchy problem (\ref{vdh})-(\ref{vdhi}) of the dual quantum Vlasov hierarchy
for the additive-type marginal observables. In case of the $k$-ary  marginal observables
a solution of the dual quantum Vlasov hierarchy (\ref{vdh}) is equivalent in the sense of equality (\ref{dchaos})
to preserving of a chaos property for $k$-particle marginal density operators.

\subsection{The dual mean-field dynamics and the nonlinear Schr\"{o}dinger equation}

If the initial state is a pure state, i.e.  $f_{1}^0=|\psi_{0}\rangle\langle\psi_{0}|$,
the Heisenberg picture of evolution of quantum many-particle systems described by
the dual quantum Vlasov hierarchy (\ref{vdh}) is an equivalent to the Schr\"{o}dinger
picture of evolution governed by the Hartree equation.

Indeed, for a system in the pure state, i.e.  $f_{1}(t)=|\psi_{t}\rangle\langle\psi_{t}|$ or in terms of the kernel
$f_{1}(t,q,q')=\psi(t,q)\psi(t,q')$ of the marginal operator $f_{1}(t)$, Vlasov kinetic equation (\ref{Vlasov1})
is transformed to the Hartree equation
\begin{equation*}\label{Hartree}
     i\frac{\partial}{\partial t} \psi(t,q) = -\frac{1}{2}\Delta_{q}\psi(t,q)+ \int dq'\Phi(q-q')|\psi(t,q')|^{2}\psi(t,q).
\end{equation*}
If the kernel of the interaction potential $\Phi(q)=\delta(q)$ is the Dirac measure
we derive the cubic nonlinear Schr\"{o}dinger equation
\begin{equation*}\label{Schr}
     i\frac{\partial}{\partial t} \psi(t,q)=-\frac{1}{2}\Delta_{q}\psi(t,q)+ |\psi(t,q)|^{2}\psi(t,q).
\end{equation*}

\subsection{On the generalized quantum kinetic equation}

We consider the relations of the evolution of observables and the evolution of
quantum states described in terms of a one-particle marginal density operator in the general case.
In case of initial states specified by a one-particle marginal density operator,
the dual BBGKY hierarchy describes the dual picture of
evolution to the picture of the evolution of states governed by the generalized quantum kinetic equation
and an infinite sequence of explicitly defined functionals of the solution of such evolution equation.

In fact, the following equality is true
\begin{equation}\label{w}
    \big\langle G(t)\big|F^c\big\rangle =\big\langle G(0)\big|F(t\mid F_{1}(t))\big\rangle,
\end{equation}
where the initial state $F^c$ is defined by (\ref{h2}) and
$F(t\mid F_{1}(t))=(F_1(t),F_2(t\mid F_{1}(t)),\ldots,$ $F_s(t;1,\ldots,s\mid F_{1}(t))$ is a sequence
of marginal functionals of the state. The functionals $F_{s}\big(t,1,\ldots,s \mid F_{1}(t)\big),\,s\geq 2$, are represented
by the expansions over products of the one-particle density operator $F_{1}(t)$
\begin{equation}\label{f}
     F_{s}\big(t,Y\mid F_{1}(t)\big)
     \doteq\sum _{n=0}^{\infty }\frac{1}{n!}\,\mathrm{Tr}_{s+1,\ldots,{s+n}}\,
     \mathfrak{V}_{1+n}\big(t,\{Y\},s+1,\ldots,s+n\big)\prod _{i=1}^{s+n} F_{1}(t,i),
\end{equation}
where the $(n+1)$-order evolution operator $\mathfrak{V}_{1+n}(t),\,n\geq0$, are defined as follows
\begin{equation}\label{skrr}
\begin{split}
  \mathfrak{V}_{1+n}&(t,\{Y\},X\setminus Y )\\
    &\doteq\sum_{k=0}^{n}\,(-1)^k\,\sum_{n_1=1}^{n} \ldots
        \sum_{n_k=1}^{n-n_1-\ldots-n_{k-1}}\frac{n!}{(n-n_1-\ldots-n_k)!}\,\widehat{\mathfrak{A}}_{1+n-n_1-\ldots-n_k}(t,\{Y\},\\
    &\hskip+15mm s+1,\ldots,s+n-n_1-\ldots-n_k)
        \prod_{j=1}^k\,\sum\limits_{\mbox{\scriptsize $\begin{array}{c}\mathrm{D}_{j}:Z_j=\bigcup_{l_j} X_{l_j},\\
        |\mathrm{D}_{j}|\leq s+n-n_1-\dots-n_j\end{array}$}}\frac{1}{|\mathrm{D}_{j}|!}\\
    &\hskip+35mm \times\sum_{i_1\neq\ldots\neq i_{|\mathrm{D}_{j}|}=1}^{s+n-n_1-\ldots-n_j}\,\,
        \prod_{X_{l_j}\subset \mathrm{D}_{j}}\,\frac{1}{|X_{l_j}|!}\,\,
        \widehat{\mathfrak{A}}_{1+|X_{l_j}|}(t,i_{l_j},X_{l_j}),
\end{split}
\end{equation}
and $\sum_{\mathrm{D}_{j}:Z_j=\bigcup_{l_j} X_{l_j}}$ is the sum over all possible
dissections $\mathrm{D}_{j}$ of the linearly ordered set
$Z_j\equiv(s+n-n_1-\ldots-n_j+1,\ldots,s+n-n_1-\ldots-n_{j-1})$ on no more than $s+n-n_1-\ldots-n_j$ linearly ordered subsets \cite{GT}.
In expression (\ref{skrr}) it is denoted by $\widehat{\mathfrak{A}}_{1+n}(t)$ the $(1+n)$-order cumulant
\begin{equation}\label{cd}
\begin{split}
    \widehat{\mathfrak{A}}_{1+n}(t,\{Y\},s+1,&\ldots,s+n)\\
    & \doteq\sum\limits_{\mathrm{P}:\,(\{Y\},s+1,\ldots,s+n)={\bigcup\limits}_i X_i}
      (-1)^{|\mathrm{P}|-1}(|\mathrm{P}|-1)!\prod_{X_i\subset \mathrm{P}}\widehat{\mathcal{G}}_{|X_i|}(t,X_i),
\end{split}
\end{equation}
of the groups of scattering operators
\begin{equation*}\label{so}
     \widehat{\mathcal{G}}_{|X_i|}(t)\equiv\mathcal{G}_{|X_i|}(-t,X_i)
     \prod _{i\in X_i}\mathcal{G}_{1}(t,i),
\end{equation*}
where $\sum_\mathrm{P}$ is the sum over all possible partitions of the set $(\{Y\},s+1,\ldots,s+n)$
into $|\mathrm{P}|$ nonempty mutually disjoint subsets $ X_i\subset(\{Y\},s+1,\ldots,s+n)$
and the group of operators $\mathcal{G}_{|X_i|}(-t)$ is adjoint to the group $\mathcal{G}_{|X_i|}(t)$
in the sense of functional (\ref{avmar-1}).

The one-particle density operator $F_{1}(t)$ is determined by the following series
\begin{equation}\label{ske}
    F_{1}(t,1)= \sum\limits_{n=0}^{\infty}\frac{1}{n!}\,\mathrm{Tr}_{2,\ldots,{1+n}}\,\,
        \mathfrak{A}_{1+n}(-t,1,\ldots,n+1)\prod _{i=1}^{n+1}F_{1}^0(i),
\end{equation}
where the cumulants $\mathfrak{A}_{1+n}(-t),\, n\geq0,$ are defined by the formula similar to (\ref{cumulant})
\begin{equation*}
  \mathfrak{A}_{1+n}(-t,1,\ldots,n+1)
  \doteq\sum\limits_{\mathrm{P}:\,(1,\ldots,n+1)={\bigcup\limits}_i X_i}
      (-1)^{|\mathrm{P}|-1}(|\mathrm{P}|-1)!\prod_{X_i\subset \mathrm{P}}\mathcal{G}_{|X_i|}(-t,X_i),
\end{equation*}
where $\sum_\mathrm{P}$ is the sum over all possible partitions of the set $(1,\ldots,n+1)$
into $|\mathrm{P}|$ nonempty mutually disjoint subsets $ X_i\subset(1,\ldots,n+1)$.

The one-particle density operator (\ref{ske}) is a solution of the following initial-value problem
\begin{equation}\label{gke}
\begin{split}
  \frac{d}{dt}F_{1}(t,1)=&-\mathcal{N}_{0}(1)F_{1}(t,1)+\mathrm{Tr}_{2}\big(-\mathcal{N}_{\mathrm{int}}(1,2)\big)\\
        &\times\sum\limits_{n=0}^{\infty}\frac{1}{n!}\mathrm{Tr}_{3,\ldots,n+2}
        \mathfrak{V}_{1+n}\big(t,\{1,2\},3,\ldots,n+2\big)\prod _{i=1}^{n+2} F_{1}(t,i),
\end{split}
\end{equation}
\begin{equation}\label{2}
   \hskip-80mm F_1(t,1)|_{t=0}= F_1^0(1),
\end{equation}
where the evolution operator $\mathfrak{V}_{1+n}(t)$ is defined by formula (\ref{skrr}). We refer to evolution equation
(\ref{gke}) as the generalized quantum kinetic equation. For systems of classical particles such equation was formulated
in \cite{CGP97} and for quantum many-particle systems in \cite{GT}.

To verify equality (\ref{w}) we transform functional $\langle G(t)|F^c\rangle$ as follows
\begin{equation}\label{tf}
\begin{split}
  & \big\langle G(t)\big|F^c\big\rangle=
     \sum_{s=0}^{\infty}\,\frac{1}{s!}\,\mathrm{Tr}_{\mathrm{1,\ldots,s}}G_{s}(0,1,\ldots,s)\\
  &\hskip+21mm \times\sum\limits_{n=0}^{\infty}\frac{1}{n!}
        \,\mathrm{Tr}_{s+1,\ldots, s+n}\,\mathfrak{A}_{1+n}(-t,\{Y\},s+1,\ldots, s+n)\prod _{i=1}^{s}F_{1}^0(i),
\end{split}
\end{equation}
where the $(1+n)$-order cumulant $\mathfrak{A}_{1+n}(-t,\{Y\},s+1,\ldots, s+n)$ is defined by (\ref{cd}).
For $F_1^{0}\in\mathfrak{L}^{1}(\mathcal{H})$ and $G_{s}(0)\in\mathfrak{L}(\mathcal{H}_s)$
obtained functional (\ref{tf}) exists under the condition $\|F_1^0\|_{\mathfrak{L}^{1}(\mathcal{H})}<e^{-1}$.

Then we expand the cumulants $\mathfrak{A}_{1+n}(-t)$ over the new evolution operators $\mathfrak{V}_{1+n}(t),\,n\geq0,$
into the kinetic cluster expansion \cite{GT}
\begin{equation*}\label{rrrl2}
\begin{split}
   &\mathfrak{A}_{1+n}(-t,\{Y\},s+1,\ldots,s+n)=\sum_{n_1=0}^{n}\frac{n!}{(n-n_1)!}\,\mathfrak{V}_{1+n-n_1}\big(t,\{Y\},s+1,\ldots,s+n-n_1\big)\\
   &\hskip+5mm\times\sum\limits_{\mbox{\scriptsize $\begin{array}{c}\mathrm{D}:Z=\bigcup_k X_k,\\|\mathrm{D}|\leq s+n-n_1\end{array}$}}
     \frac{1}{|\mathrm{D}|!}\,\sum_{i_1\neq\ldots\neq i_{|\mathrm{D}|}=1}^{s+n-n_1}\,
     \prod_{X_{k}\subset \mathrm{D}}\,\frac{1}{|X_k|!}\,\mathfrak{A}_{1+|X_{k}|}(-t,i_k,X_{k})
     \prod\limits_{\mbox{\scriptsize$\begin{array}{c}{m=1},
     \\m\neq i_1,\ldots,i_{|\mathrm{D}|}\end{array}$}}^{s+n-n_1}\mathfrak{A}_1(-t,m),
\end{split}
\end{equation*}
where $\sum_{\mathrm{D}:Z=\bigcup_l X_l,\,|\mathrm{D}|\leq s+n-n_1}$ is the sum over all possible dissections
$\mathrm{D}$ of the linearly ordered set $Z\equiv(s+n-n_1+1,\ldots,s+n)$ on no more than $s+n-n_1$ linearly ordered subsets.
Representing series over the summation index $n$ and the sum over the summation index $n_1$ in functional (\ref{tf})
as the two-fold series and identifying the series over the summation index $n_1$ with the products
of one-particle density operators similar to formula (\ref{prodi})
\begin{equation*}
\begin{split}
  &\sum_{n_1=0}^{\infty}\mathrm{Tr}_{s+n+1,\ldots,s+n+n_1}
     \sum\limits_{\mbox{\scriptsize $\begin{array}{c}\mathrm{D}:Z=\bigcup_k X_k,\\|\mathrm{D}|\leq s+n\end{array}$}}
     \,\,\sum_{i_1<\ldots<i_{|\mathrm{D}|}=1}^{s+n}\,\prod_{X_{k}\subset \mathrm{D}}\,\frac{1}{|X_k|!}\\
  &\hskip+12mm\times\mathfrak{A}_{1+|X_{k}|}(-t,i_k,X_{k})\prod\limits_{\mbox{\scriptsize$\begin{array}{c}{l=1},
     \\l\neq i_1,\ldots, i_{|\mathrm{D}|}\end{array}$}}^{s+n}
     \mathfrak{A}_1(-t,l)\prod_{j=1}^{n+s+n_1}F_1^0(j)=\prod _{i=1}^{s+n} F_{1}(t,i),
\end{split}
\end{equation*}
we transform functional (\ref{tf}) to the form in terms of marginal functionals of the state (\ref{f}).
Thus, equality (\ref{w}) holds.

In a particular case of the additive-type marginal observables $G^{(1)}(0)$
equality (\ref{w}) is reduced to the form
\begin{equation*}\label{rw}
    \big\langle G^{(1)}(t)\big|F^c\big\rangle = \mathrm{Tr}_{\mathrm{1}}\,G^{(1)}(0,1)F_{1}(t,1),
\end{equation*}
where the one-particle marginal density operator $F_{1}(t)$ is a solution of the Cauchy problem (\ref{gke})-(\ref{2}).
Hence for additive-type marginal observables the generalized quantum kinetic equation (\ref{gke})
is dual to the dual quantum BBGKY hierarchy (\ref{dh}) with respect to bilinear form (\ref{avmar-1}).

Thus, in case of initial data (\ref{h2}) which is completely characterized
by the one-particle marginal density operator $F_{1}^0$, solution (\ref{sdh})
of the Cauchy problem (\ref{dh})-(\ref{dhi}) of the dual quantum BBGKY hierarchy
for marginal observables and a solution of the Cauchy problem of the generalized
kinetic equation (\ref{gke})-(\ref{2}) together with marginal functionals of the state (\ref{f})
give two equivalent approaches to the description of the evolution of quantum many-particle systems.

\section{Conclusions}
We develop an approach of the description of kinetic evolution of quantum many-particle systems in terms
of the evolution of marginal observables. One of the advantage of such approach is the possibility
to construct the kinetic equations in scaling limits if there
are correlations of particle states at initial time \cite{CGP97}, for instance,
correlations characterizing the condensate states \cite{BQ}.

In the case of quantum systems of particles obeying Fermi or Bose statistics \cite{GP}
quantum kinetic equations have the different structure from formulated above.
The analysis of these cases will be given in a separate paper.

Finally we point out the relation of the generalized quantum kinetic equation (\ref{gke})
and the specific quantum kinetic equations. The last can be derived
from the generalized quantum kinetic equation in the appropriate scaling limits \cite{Sh} or as a result of certain
approximations. For example, in the mean-field limit we derive the quantum Vlasov kinetic equation \cite{GT}.
Observing that in the kinetic (macroscopic) scale of the variation of variables \cite{CIP} the groups of operators (\ref{grG})
of finitely many particles depend on microscopic time variable $\varepsilon^{-1}t$, where $\varepsilon\geq0$ is a scale
parameter, the dimensionless marginal functionals of the state are represented in the form:
$F_{s}\big(\varepsilon^{-1}t,Y\mid F_{1}(t)\big)$. Then in the limit $\varepsilon\rightarrow0$ the first two terms
of the dimensionless marginal functional expansions (\ref{f})
\begin{equation*}
\begin{split}
    & \widehat{\mathcal{G}}_{s}(\varepsilon^{-1}t,Y)\prod _{i=1}^{s} F_{1}(t,i)\\
    &\hskip+5mm +\int_0^{\varepsilon^{-1}t} d\tau\,\mathcal{G}_{s}(-\tau,Y)\mathrm{Tr}_{s+1}\big(\sum\limits_{i_1=1}^{s}
       (-\mathcal{N}_{\mathrm{int}}(i_1,s+1))\widehat{\mathcal{G}}_{s+1}(\varepsilon^{-1}t,Y,s+1)\\
    &\hskip+12mm  -\widehat{\mathcal{G}}_{s}(\varepsilon^{-1}t,Y)\sum\limits_{i_1=1}^{s}(-\mathcal{N}_{\mathrm{int}}(i_1,s+1))
       \widehat{\mathcal{G}}_{2}(\varepsilon^{-1}t,i_1,s+1)\big)\prod_{i_2=1}^{s+1}\mathcal{G}_{1}(\tau,i_2)F_{1}(t,i_2)
\end{split}
\end{equation*}
coincide with corresponding terms constructed by the perturbation
method with the use of the weakening of correlation condition by Bogolyubov \cite{CGP97}
Thus, in the kinetic scale the collision integral of the generalized kinetic equation (\ref{gke}) takes the form of
Bogolyubov's collision integral \cite{CGP97},\cite{Ger} which enables to control correlations
of infinite-particle systems. In a space homogeneous case the collision integral of the first approximation has
a more general form than the quantum Boltzmann collision integral.

\end{document}